# Entanglement in Quantum Dots: Insights from Dynamic Susceptibility and Quantum Fisher Information

Jahanfar Abouie[a*] and Daryoosh Vashaee[b,c*]

[a] Department of Physics, Institute for Advanced Studies in Basic Sciences, Zanjan 45137-66731, Iran

[b] Department of Materials Science & Engineering, North Carolina State University, Raleigh, NC 27606, USA

[c] Department of Electrical & Computer Engineering, North Carolina State University, Raleigh, NC 27606, USA

**Abstract**

This study investigates the entanglement properties of quantum dots (QDs) under a universal Hamiltonian where the Coulomb interaction between particles (electrons or holes) decouples into charging energy and exchange coupling terms. Although this formalism typically decouples the charge and spin components, confinement-induced energy splitting can induce unexpected entanglement within the system. By analyzing the dynamic susceptibility and quantum Fisher information (QFI), we uncover significant behaviors influenced by exchange constants, temperature variations, and confinement effects. In QDs with Ising exchange interactions, far below the Stoner instability (SI) point, where the QD is in a disordered paramagnetic phase, temperature reductions lead to decreased entanglement, challenging conventional expectations. Our findings demonstrate that for QDs with small exchange interactions, the responses of easy-plane ($J_z < J_\perp$) and easy-axis ($J_z > J_\perp$) configurations are similar, with increased anisotropy broadening susceptibility and shifting its maximum to higher frequencies. For large exchange interactions, the susceptibility differences between easy-plane and easy-axis QDs become significant, with easy-plane QDs exhibiting a higher susceptibility magnitude. Additionally, the study reveals that temperature variations affect the dynamic response functions differently in easy-axis and easy-plane QDs. In easy-plane QDs, QFI consistently decreases with increasing temperature, whereas in easy-axis QDs, QFI behavior is highly dependent on the strengths of $J_z$ and $J_\perp$, showing either an increase or decrease with temperature based on specific coupling conditions. Conversely, at low temperatures, anisotropic Heisenberg models exhibit enhanced entanglement near isotropic points. Overall, this work contributes to advancing our understanding of entanglement in QDs and its potential applications in quantum technologies.

## Introduction

Quantum entanglement, a fundamental concept in quantum physics, serves as a key to unlocking the mysteries of correlations that extend beyond classical physics, propelling advancements in quantum technologies. Exploring entanglement not only enriches our understanding of quantum many-body physics but also drives the development of experimental protocols aimed at detecting and quantifying entanglement in many-body quantum states. Such endeavors are crucial for leveraging entanglement as a resource in quantum computing, communication, and metrology.

* Corresponding authors: jahan@iasbs.ac.ir and dvashae@ncsu.edu

Recent studies have made significant strides by introducing a method to measure multipartite entanglement in thermal ensembles through dynamic susceptibility.[1] This technique finds immediate applicability with existing experimental setups in cold atomic gases and condensed matter systems.[2,3] By enabling direct measurement of multipartite entanglement, essential for advancing quantum technologies, this approach also establishes a vital link between multipartite entanglement and observable many-body correlations in experiments.

The elusive nature of entanglement in large quantum systems poses a challenge for its direct measurement, particularly in characterizing quantum phases and transitions. Traditional methods like tomography face limitations due to exponential resource scaling, making them impractical for systems beyond a few qubits.[4,5] In contrast, quantum Fisher information (QFI), a concept from quantum metrology, can serve as an indicator for multipartite entanglement. QFI, which quantifies the maximum precision achievable in parameter estimation, becomes directly measurable through dynamic susceptibilities.[5,6] This approach enables the direct measurement of multipartite entanglement in quantum systems.

Quantum dots (QDs), with their discrete energy levels and tunable properties, offer a fertile ground for exploring quantum phenomena at the nanoscale. Among these phenomena, the interplay between quantum entanglement and exchange interactions, both anisotropic Ising and isotropic Heisenberg, presents a rich area of study, especially concerning the Stoner instability (SI). Understanding entanglement between spin and charge degrees of freedom in QDs is vital for advancing quantum information technologies.

The exchange interactions within QDs exhibit notable distinctions in the presence and absence of spin-orbit coupling (SOC), with Ising exchange predominating in the presence of SOC and Heisenberg exchange prevailing in its absence.[7] Additionally, it is essential to recognize that holes and electrons behave distinctively in QDs due to their differing interactions with SOC: holes are typically influenced by SOC, whereas electrons are not. This differentiation underscores the significance of comprehending the unique behaviors of electrons and holes in QDs, thereby enriching our insight into spin-charge entanglement within these systems.

Entanglement, a fundamental concept in quantum physics whereby particles become interconnected to the extent that the overall state of the system cannot be expressed as the product of individual particle states, serves as an essential resource for quantum computation, communication, and metrology. Notably, the capability to generate and manipulate entangled states among various types of qubits, such as spin and charge qubits, paves the way for novel quantum operations and augments the functionality of quantum devices.

The Stoner instability, a quantum phase transition from a paramagnetic to a ferromagnetic state, is a cornerstone concept for understanding magnetism in QDs, with practical implications for their electronic and magnetic properties. The tunneling density of states in QDs and nanoparticles has been reevaluated with the extension of the universal Hamiltonian to include uniaxial anisotropic exchange, shedding light on the systems' behavior near the SI point. The significance of the ground state's total spin in determining the tunneling density of states was highlighted, revealing a single maximum in its energy dependence, which contrasts with previous perturbative analyses predicting multiple extrema.[7]

Dynamic (or AC) susceptibility measurements offer a non-invasive means to probe the dynamic response of a system to an external alternating magnetic field, providing insights into intrinsic properties like magnetic susceptibility and dielectric permittivity. In the context of quantum dots, dynamic susceptibility serves as a proxy to analyze the intricate dynamics of spin and charge entanglement.[1,8] By examining the response of quantum dots to varying frequencies and amplitudes of an AC field, researchers can infer the degree of entanglement and coherence between qubits, essential for implementing quantum gates and algorithms.

This paper highlights the methodological importance within QD systems, holding significant experimental implications. It provides a comprehensive analysis of spin-exchange interactions within QDs, contributing to a broader understanding of charge-spin entanglement in these systems. Various exchange interactions, including Ising and Heisenberg models, are considered, covering both isotropic and anisotropic scenarios, with analyses spanning both high and low-temperature extremes. Such comparisons deepen understanding of the mesoscopic SI and the effects of exchange interactions on the quantum states of QDs, particularly in the context of entanglement, where exchange interactions play a crucial role in entangling the spins of confined electrons or holes within QDs. The distinction between Heisenberg and Ising exchange interactions introduces complexity in understanding the quantum behavior of QDs, especially with the presence of SOC, which can significantly affect the quantum states and entanglement properties of the system.

The study employs the universal Hamiltonian approach to simplify the intricate electron-electron interactions within QDs, focusing on zero-mode interactions that encompass charging and spin-exchange terms. While this formalism effectively decouples the charge and spin components, our analysis reveals that the confinement-induced energy splitting unexpectedly induces entanglement between electrons' spin in the system. However, it is crucial to acknowledge that strong SOC poses a challenge to the applicability of the universal Hamiltonian to QDs. Under strong SOC conditions, fluctuations in the interaction matrix elements become significant, particularly in the metallic regime where the mean-level spacing ($\Delta$) relative to the Thouless energy ($E_{Th}$) is very small (i.e., $\Delta/E_{Th} \ll 1$). This complexity suggests that the universal Hamiltonian may not fully capture the behavior of QDs in the presence of strong SOC, indicating the need for a more detailed approach to understand these systems.[9,10] Furthermore, it is worth noting that this regime is not suitable for a single QD qubit characterized by one or a small number of electrons. Additionally, the paper explores the potential experimental determination of entanglement, a critical parameter for understanding and manipulating QD dynamics.

By comparing Ising and Heisenberg exchange interactions, both anisotropic and isotropic, the study provides a clearer picture of entanglements in each case, advancing theoretical models and suggesting experimental pathways for probing the intricate dynamics of charge and spin within quantum dots. These findings have implications for the development of quantum computing and spintronic devices, where precise control and understanding of spin and charge are essential. The linkage between QFI and dynamic susceptibilities not only facilitates experimental identification of entanglement but also deepens theoretical insight into quantum fluctuations and their influence on the sensitivity of quantum states to external disturbances.

## Fundamental Formalism

The anisotropic exchange constant and non-zero dynamical susceptibility are essential parameters in characterizing QD systems. The anisotropic exchange constant describes the interactions among electron spins within QDs, which vary based on their relative orientations. This parameter significantly influences spin dynamics within the QD, playing a key role in quantum information processing applications.

Similarly, the non-zero dynamical susceptibility reveals how the QD responds to an externally applied field, with its sensitivity dependent on the field's frequency. Understanding this parameter is crucial for devising effective strategies in device engineering and manipulating quantum states within the system.

Magnetic susceptibility, a fundamental property of systems, measures their responsiveness to applied magnetic fields, typically manifested through magnetization fluctuations. This property is intimately linked to spontaneous magnetization within the system.

In paramagnetic systems, devoid of spontaneous magnetization, magnetic susceptibility denotes the magnetization induced by an external magnetic field, expressed as the proportionality constant relating induced magnetization (M) to applied magnetic field strength (H). This relationship is often represented as M = χH, where χ represents the magnetic susceptibility.

For a time-varying external magnetic field described by $\vec{h}(t) = \vec{h}_0 e^{-i\omega t}$, where $h_0$ denotes the field magnitude, ω signifies the angular frequency, and 'i' denotes the imaginary unit, the system's response evolves temporally. Consequently, the concept of dynamical susceptibility emerges, capturing the system's response as a function of the frequency (ω) of external fields.

Understanding magnetic susceptibility is crucial in studying quantum systems like QDs, especially concerning gate-defined QD qubits. In these systems, magnetization ($M_\alpha(t)$) is fundamental and is expressed as:

$$M_\alpha(t) = \langle S_\alpha(t) \rangle = \chi_{\alpha\beta} h_\beta(t). \quad (1)$$

Here, the indices α and β represent the x, y, z dimensions, and magnetization corresponds to the correlation function:

$$M_\alpha(t) = -i \int_{-\infty}^{t} dt' \langle [S_\alpha(t), V(t')] \rangle. \quad (2)$$

In this equation, $V = -\vec{S}.\vec{h}(t)$ represents the Hamiltonian governing the interaction of electron spin with the external magnetic field. By utilizing equations (1) and (2), the magnetic susceptibility can be expressed as:[11]

$$\chi_{\alpha\beta}(\omega) = -i \int_{-\infty}^{t} dt' e^{i\omega(t-t\prime)} \langle [S_\alpha(t), S_\beta(t')] \rangle. \quad (3)$$

Equation (3) denotes the Fourier transform of a retarded correlation function. Retarded correlation functions can be derived from Matsubara functions via the analytic continuation $i\omega_n \longrightarrow \omega + i\delta$, where δ is infinitesimal. Expressed in terms of the Matsubara correlation function, the magnetic susceptibility becomes:

$$\chi_{\alpha\beta}(i\omega_n) = \int_0^\beta d\tau e^{i\omega_n \tau} \langle T_\tau S_\alpha(t) S_\beta(0) \rangle. \quad (4)$$

Here, $\omega_n = (2n+1)\pi k_B T$, where T represents temperature and constitutes the poles of the fermionic distribution function. $T_\tau$ is the time-ordering operator.

Furthermore, the correlation function can be evaluated by defining the wave-vector dependent

susceptibility:[11]

$$\chi_{\alpha\beta}(\vec{q},i\omega_n)=\int_0^{\beta} d\tau e^{i\omega_n\tau}\langle T_\tau S_\alpha(\vec{q},\tau)S_\beta(-\vec{q},0)\rangle. \quad (5)$$

The spin operators are typically expressed in terms of electron creation and annihilation operators:

$$S_\alpha(\vec{q})=\frac{1}{2}\sum_{\vec{p}\sigma\sigma'} a^\dagger_{\vec{p}+\vec{q}\sigma}\sigma^\alpha_{\sigma\sigma'}a_{\vec{p}\sigma'}\,. \quad (6)$$

Here, $\sigma^\alpha$ ($\alpha$=x,y,z) represents the Pauli spin matrices. The dynamic longitudinal and transverse susceptibilities can be calculated as follows:[11]

$$\begin{cases}\chi_{zz}(\vec{q},i\omega_n)=\frac{1}{4}\sum_{\vec{p}\vec{k}ss'} ss'\int_0^{\beta} d\tau e^{i\omega_n\tau}\langle T_\tau a^\dagger_{\vec{p}+\vec{q}s}(\tau)a_{\vec{p}s}(\tau)a^\dagger_{\vec{k}-\vec{q}s'}(0)a_{\vec{k}s'}(0)\rangle\,,\\ \chi_{+-}(\vec{q},i\omega_n)=\sum_{\vec{p}\vec{k}}\int_0^{\beta} d\tau e^{i\omega_n\tau}\langle T_\tau a^\dagger_{\vec{p}+\vec{q}\uparrow}(\tau)a_{\vec{p}\downarrow}(\tau)a^\dagger_{\vec{k}-\vec{q}\downarrow}(0)a_{\vec{k}\uparrow}(0)\rangle\,,\\ \chi_{-+}(\vec{q},i\omega_n)=\sum_{\vec{p}\vec{k}}\int_0^{\beta} d\tau e^{i\omega_n\tau}\langle T_\tau a^\dagger_{\vec{p}+\vec{q}\downarrow}(\tau)a_{\vec{p}\uparrow}(\tau)a^\dagger_{\vec{k}-\vec{q}\uparrow}(0)a_{\vec{k}\downarrow}(0)\rangle\,.\end{cases} \quad (7)$$

The crucial step in calculating the magnetic susceptibility is to derive an expression for the bracket in equation (7). For a noninteracting free electron gas, these operators can be paired using Wick's theorem. To obtain both the dynamic and static susceptibilities, a single-particle Green's function is defined as:

$$G^{(0)}_s(\vec{p},\tau)=-\langle T_\tau a_{\vec{p}s}(\tau)a^\dagger_{\vec{p}s}(0)\rangle, \quad (8)$$

where $s=\uparrow,\downarrow$. This function can be expanded into a Fourier series:

$$G^{(0)}_s(\vec{p},\tau)=\frac{1}{\beta}\sum_n e^{-i\omega_n\tau}G^{(0)}_s(\vec{p},i\omega_n). \quad (9)$$

Here, $G^{(0)}_s(\vec{p},i\omega_n)=\left(i\omega_n-\xi_{\vec{p}}\right)^{-1}$ with $\xi_{\vec{p}}=\varepsilon_{\vec{p}}-\mu$, where $\varepsilon_{\vec{p}}$ is the single-particle energy and $\mu$ is the chemical potential. In terms of single-particle Green's function the transverse dynamic susceptibility of a noninteracting electron gas is given by:[11]

$$\chi^{(0)}_{+-}(\vec{q},i\omega_n)=\int_0^{\beta} d\tau e^{i\omega_n\tau}\sum_{\vec{k}} G^{(0)}_\uparrow\left(\vec{k}+\vec{q},-\tau\right)G^{(0)}_\downarrow\left(\vec{k},\tau\right)=\frac{1}{N}\sum_{\vec{k}}\frac{n_{\mathrm{F}}\left(\xi_{\vec{k}}\right)-n_{\mathrm{F}}\left(\xi_{\vec{k}+\vec{q}}\right)}{i\omega_n+\xi_{\vec{k}}-\xi_{\vec{k}+\vec{q}}}, \quad (10)$$

where $n_{\mathrm{F}}$ is the Fermi distribution function, and the longitudinal susceptibility is given in terms of transverse as $\chi^{(0)}_{+-}(\vec{q},i\omega_n)=2\chi^{(0)}_{zz}(\vec{q},i\omega_n)$.

The static susceptibility is given by $\chi=\lim_{\vec{q}\to 0}\chi_{zz}(\vec{q},i\omega_n=0)$. For noninteracting systems, the static susceptibility reduces to $\chi^{(0)}=\mu_{\mathrm{B}}^2 N_{\mathrm{F}}$, where $\mu_{\mathrm{B}}$ is the Bohr magneton and $N_{\mathrm{F}}$ is the electron density of state at the Fermi energy.

**Electron-Electron Interaction and Universal Hamiltonian in QDs**

In the metallic regime characterized by high Thouless conductance ($g_T\gg 1$), the behavior of QDs is fully described by a universal Hamiltonian. This Hamiltonian encapsulates the electron-electron interaction within the QD and comprises three spatially independent components: the charging energy, the ferromagnetic exchange coupling, and the Cooper channel interaction.[12] In the presence of the first two terms, the universal Hamiltonian for QDs is expressed as follows:

$$H=H_0+H_C+H_S\,,$$

$$H_0 = \sum_{\alpha\sigma} \epsilon_\alpha a^\dagger_{\alpha\sigma} a_{\alpha\sigma}\,,\;\; H_C = E_C(\hat{n} - N_0)^2,\;\; H_S = -J_\perp\left(S_x^2 + S_y^2\right) - J_z S_z^2\,. \quad (11)$$

Here, $H_0$ represents the non-interacting Hamiltonian, where $a^\dagger_{\alpha\sigma}$ is the electron creation operator for an orbital level α and spin $\frac{\sigma}{2}\,(\sigma = \pm 1)$, and $\epsilon_\alpha$ denotes spin-degenerate single-particle levels. In a general case, the orbital energies $\epsilon_\alpha$ are random, and the mean-level spacing Δ is defined as $\langle \epsilon_{\alpha+1} - \epsilon_\alpha \rangle$, where $\langle \cdots \rangle$ indicates ensemble averaging. $H_C$ stands for the charging interaction, with $\hat{n} = \sum_\alpha \hat{n}_\alpha = \sum_{\alpha,\,\sigma=\uparrow,\downarrow} a^+_{\alpha\sigma} a_{\alpha\sigma}$ representing the number operator. $E_C$ represents the charging energy, and $N_0$ represents a positive background charge. The last term in equation (11), where $S_{i(=x,y,z)} = \sum_\alpha S_{i\alpha},\; S_{i\alpha} = \frac{1}{2}\sum_{\sigma\sigma'} a^+_{\alpha\sigma} \boldsymbol{\sigma}^i_{\sigma\sigma'} a_{\alpha\sigma'}$, encapsulates exchange interactions within the QD, with $\sigma$ representing the Pauli vector matrix, and $J_z$ and $J_\perp$ denoting the exchange interaction constants.

It is well-recognized that in the presence of ferromagnetic exchange interaction, bulk systems may undergo a Stoner transition from a paramagnetic to a ferromagnetic phase.[12] However, in finite systems like QDs, an intermediate phase may emerge depending on the strength of the exchange constants, sometimes between the paramagnetic and ferromagnetic phases. In QDs with isotropic Heisenberg exchange constant (i.e., when $J_z = J_\perp = J$), three different regimes are observed:[12]

1. At small values of $J \ll \Delta$, satisfying $g_T = E_T/\Delta \gg 1$, the total spin of the QD is zero, and the QD is in the disordered paramagnetic phase.
2. As $J$ increases, the ferromagnetic exchange coupling attempts to align the spins of electrons, and the competition between the exchange coupling and the kinetic energy leads to an intermediate partially polarized phase in which the total spin of the QD becomes nonzero. This intermediate regime occurs in the interval $\Delta/2 \lesssim J < \Delta$, and the total spin of the QD increases monotonically with $J$ as $S = J/[2(\Delta - J)]$. This intermediate regime is referred to as the mesoscopic Stoner regime, which disappears as $\Delta \longrightarrow 0$.
3. By further increasing $J$ towards Δ, the system approaches the thermodynamic SI point, and the QD undergoes a second-order phase transition to a spin-polarized ferromagnetic phase. In the mesoscopic Stoner regime, the total spin is finite but not dependent on the volume of the system. However, in the ferromagnetic phase, all electrons become fully spin-polarized, and the magnetization is proportional to the volume of the QD.

For QDs with anisotropic exchange Hamiltonian ($J_z \neq J_\perp$), the mesoscopic Stoner regime emerges in the interval $\Delta - J_z \ll J_\perp < J_z$. In this regime, the total spin is approximately given by[15]: $S \cong (J_\perp + J_z - \Delta)/[\,2(\Delta - J_z)]$

However, for the case of an Ising exchange Hamiltonian, the mesoscopic Stoner regime vanishes.[12] In this scenario, if $J_z < \Delta$, the QD resides in the disordered paramagnetic phase with $S = 0$. Conversely, as $J_z$ increases towards Δ, the paramagnetic phase undergoes an abrupt transition to the fully polarized ferromagnetic phase at $J = \Delta$.

**Dynamic Transverse Susceptibility Analysis in QDs with Ising Exchange Hamiltonian**

An exact solution for QDs with Ising exchange constant $-J_z S_z^2$, ($J_\perp = 0$) has been presented in [13] in the Coulomb blockade (CB) regime. In their work, I. S. Burmistrov et al. considered the spin-disordered regime below the SI, wherein the parameters of the Hamiltonian (11) satisfy $J_z < \Delta \ll T \ll E_c$. The condition $E_c \gg \Delta$ is met in relatively large QDs, as it scales with size $L$ as $1/L^2$, while the charging energy scales as $E_c \sim 1/L$. For a two-dimensional QD, $E_c \gg \Delta$ is equivalent to $L \gg a_B$, where $a_B$ represents the effective Bohr radius for electrons in the dot.

In the Ising model, the absence of spin-flip processes results in the dynamic longitudinal susceptibility vanishing, leaving only the static component non-zero. To calculate the dynamic transverse susceptibility, as outlined in equation (7), we must evaluate the average $\langle T_\tau a^\dagger_{\vec{p}+\vec{q}\uparrow}(\tau) a_{\vec{p}\downarrow}(\tau) a^\dagger_{\vec{k}-\vec{q}\downarrow}(0) a_{\vec{k}\uparrow}(0)\rangle$ using the Hamiltonian defined by $H$. In the CB regime the dynamic transverse susceptibility in terms of imaginary time is:[14]

$$\chi^{+-}(\tau) = \frac{\beta e^{J_z\tau}}{\tilde{Z}(\mu)} \sum_N e^{-\beta E_c (N-\tilde{N}_0)^2} \sum_{M=-N}^{N} \{ e^{-\frac{\beta}{4}(\Delta - J_z)M^2} e^{J_z \tau M} \sum_\alpha [1 - n_\alpha(\bar{\mu}_\uparrow)] n_\alpha(\bar{\mu}_\downarrow) \} \quad (12)$$

where $\beta = \frac{1}{T}$ ($k_{\mathrm{B}} = 1$) is temperature inverse. In this equation, $\bar{\mu}_\sigma \equiv N_\sigma \Delta$, and $N_\sigma$ is the total number of electrons with spin σ=↑,↓. $\tilde{N}_0 \equiv N_0 + \frac{\mu}{2E_c}$ , and $\tilde{Z}(\mu)$ denotes the grand canonical partition function. The double summation above arises from replacing the grand canonical partition function in terms of the sum over canonical ones. The summation parameters are the electrons number, $N$ and the total spin of the dot (in units of $\hbar/2$), $M$.

Below the Stoner Instability (SI) point, in the disordered regime, there is no symmetry breaking in the system to distinguish opposite spin polarizations. In this regime, at high temperatures (i.e., when the QD contains many electrons such that $N\Delta \gg T$ and $(N - |M|)\Delta \gg T$, the sum over $M$ in Eq. (12) can be replaced by an integral from $-\infty$ to $+\infty$. After Fourier-transforming $\chi^{+-}(\tau)$ to Matsubara frequencies and performing analytic continuation, the imaginary part of the physical response function $\chi^{+-}(\omega)$ is acquired as:[14]

$$\mathrm{Im}\chi^{+-}(\omega) = \frac{\sqrt{\pi\beta(\Delta - J_z)}}{2J_z} e^{\frac{\beta}{4}\left[\Delta + J_z - (\Delta - J_z)\left(\frac{\omega}{J_z}\right)^2\right]} (1 + \omega/J_z) \frac{\sinh[\beta\omega/2]}{\sinh[\frac{\beta\Delta}{2}(1+\omega/J_z)]} . \quad (13)$$

The imaginary component of $\chi^{+-}(\omega)$ signifies the system's capacity to absorb and dissipate magnetic energy at a nonzero frequency, offering a vital understanding of the system's dynamics. It's evident that the transverse susceptibility remains unaffected by the charging interaction in the QD. Indeed, under the specified conditions, the charge and spin degrees of freedom are effectively decoupled.

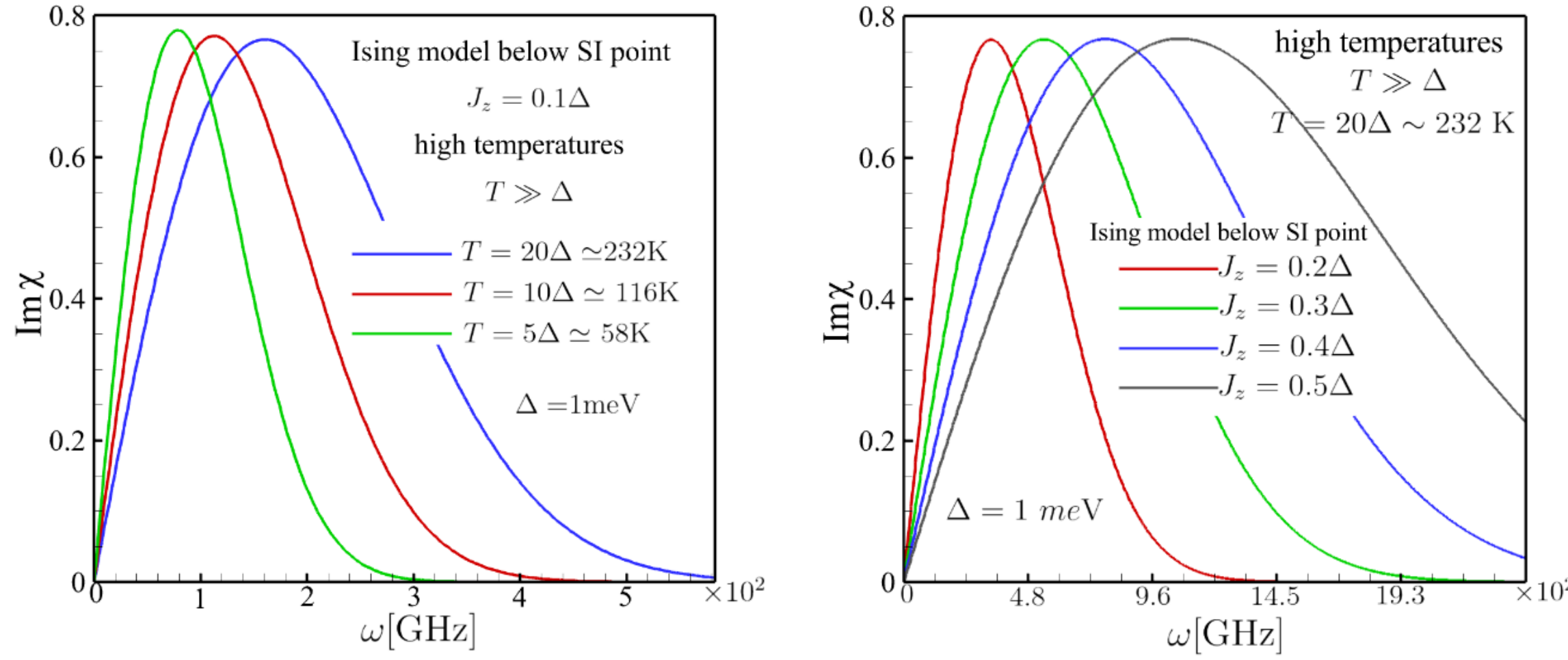

Figure 1: Comparison of the imaginary component of the dynamic transverse susceptibility of QDs with Ising exchange Hamiltonian as a function of frequency ($\omega$): depicting variations across a range of temperatures (left graph) and different values of the exchange constant (right graph).

Figure 1 illustrates the plot of the function Im$\chi^{+-}$ with respect to frequency ($\omega$). The left plot corresponds to $J_z = 0.1\Delta$ for various temperature values that satisfy $\frac{\Delta}{T} \ll 1$, while the right plot is configured at $T = 20\Delta$ for different $J_z$ values less than $\Delta$. The characteristic curve of Im$\chi^{+-}$ starts from zero at $\omega$=0, increases linearly at small $\omega$ values, reaches a maximum at $\omega_0 \approx \sqrt{\frac{2TJ^2}{(\Delta - J)}}$, and eventually decreases for larger $\omega$ values.

The location of the maximum primarily depends on the strength of the coupling constant, while its value remains unaffected. Furthermore, the peak value weakly depends on the QD temperature, but the peak location increases with the rising QD temperature. Lowering the temperature reduces the maximum magnetic energy absorbed or dissipated by the QD, primarily due to the decrease in thermal fluctuations.

The full width of susceptibility at half-maximum is directly related to the resonance frequency, as demonstrated by the relation $w$=1.59$\omega_0$. The width of the susceptibility is proportional to magnetization fluctuations. As $J_z$ increases towards $\Delta$, approaching the SI point, fluctuations in magnetization increase. Consequently, the susceptibility curve becomes broader. Moreover, decreasing the temperature reduces magnetization fluctuations, resulting in a narrower susceptibility curve.

The resonance frequency is on the order of 100 GHz, which can be challenging to access in experiments. However, by applying an external dc magnetic field, Zeeman splitting is induced in the degenerate single-electron energy levels. In this case, it is possible to observe the aforementioned nontrivial behavior of the dynamic transverse spin susceptibility at lower frequencies, accessible in research laboratories.

Regarding the static susceptibility, only the real part is finite. A straightforward calculation yields:

$$Re\chi^{+-}(0) = \frac{e^{\frac{1}{4T}(\Delta + J_z)}}{\Delta}. \quad (14)$$

It's worth noting that for $J_z = 0$, we revert to the well-known identity $\chi^{+-}(0) = 2\chi^{zz}$ for static susceptibilities. The real part of $\chi^{+-}$ at finite frequencies can be obtained either directly or through the application of Kramers-Kronig relations. However, the result is not discussed here due to its minor physical relevance.

An important point to mention is that the dynamic transverse magnetic susceptibility in Eq. (13) does not depend on the charging energy $E_C$, and is solely a function of the exchange constant $J_z$ and the mean level spacing $\Delta$. This implies that the charge-spin coupling in QD is very weak in the limit $J_z < \Delta \ll T \ll E_c$, under the conditions $N\Delta \gg 1$ and $(N - |M|)\Delta \gg 1$.

If the exchange interaction of electrons in the QD follows the isotropic Heisenberg Hamiltonian, resulting in isotropic exchange interaction ($J_z = J_\perp$), the dynamical spin response $\chi(\omega \neq 0)$ does not exist unless the dot is connected to reservoirs. Actually, the dynamical susceptibility of QDs with the isotropic Heisenberg model ($J_z = J_\perp$) is fully captured by magnetization $M$ as $\mathrm{Im}\chi(\omega) = 2\pi M\delta(\omega)$, where $\delta(\omega)$ is the Dirac delta function. Therefore, as long as the magnetization is zero, the susceptibility is also zero. As we discussed, far below the SI point, in the disordered paramagnetic phase, the system exhibits no magnetization in the absence of an external uniform magnetic field. In the presence of an external magnetic field $b$, the dynamical susceptibility is given by $2\pi M\delta(\omega - b)$, where nonzero magnetization results in a finite dynamical susceptibility.[15]

**Spin-Spin Entanglement in QDs Using Quantum Fisher Information**

The Quantum Fisher Information (QFI) acts as a quantifiable measure for multipartite entanglement, discerning the entanglement intricacies embedded within the many-body quantum correlations inherent in dynamic susceptibility. This relationship between the QFI and $\mathrm{Im}\chi$ is mathematically expressed as:[1]

$$\mathrm{QFI} = \frac{4}{\pi}\int_0^\infty d\omega \tanh\left[\frac{\beta\omega}{2}\right] \mathrm{Im}\chi. \quad (15)$$

By substituting equation (13) into integral (15), we can effectively derive the QFI, thereby illuminating the entanglement originating from dynamic transverse spin correlations within the QD.

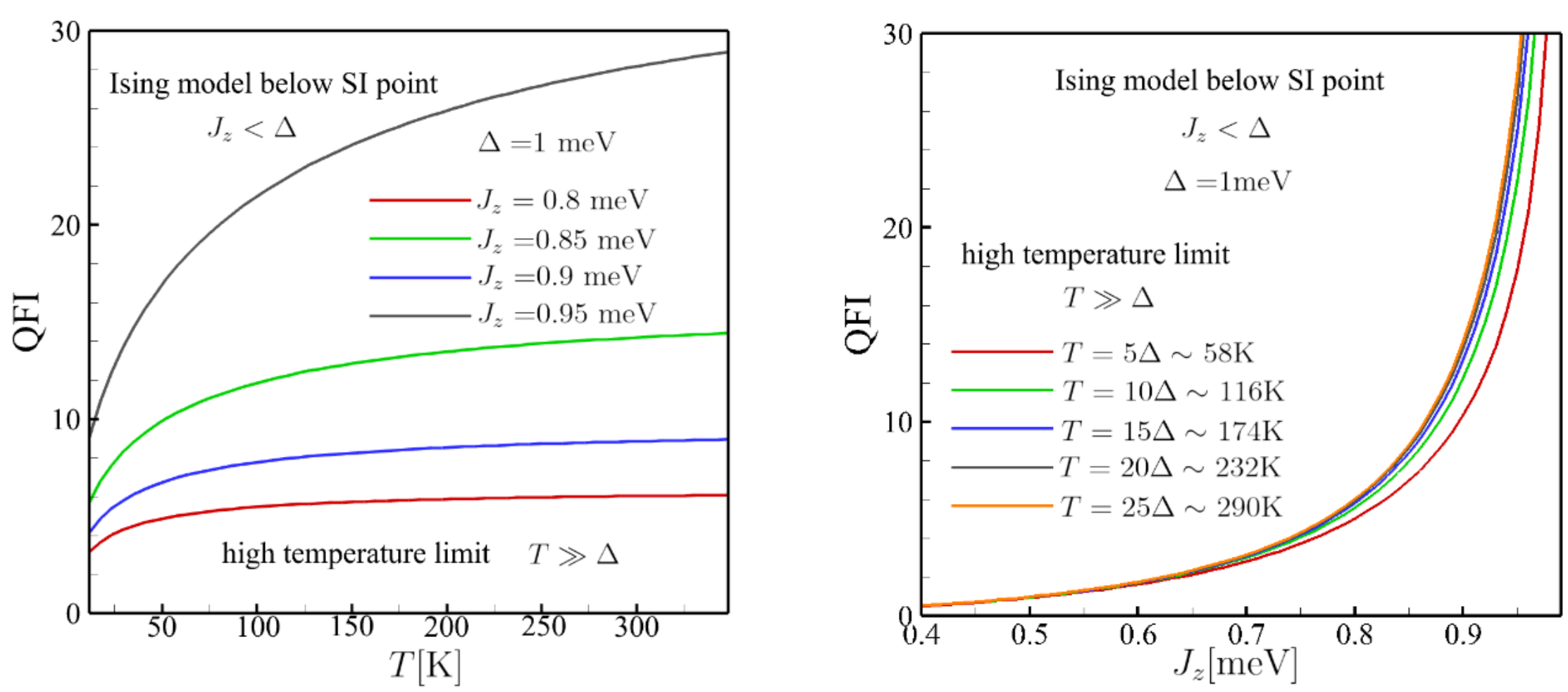

Figure 2: Quantum Fisher Information (QFI) in QDs with Ising exchange Hamiltonian plotted as a function of temperature for various exchange constant strengths, situated far from the SI point (left), and as a function of the exchange interaction constant $J_z$ across different temperatures (right).

Figure 2 depicts the QFI as a function of temperature for various strengths of the exchange constant (on the left) and as a function of $J_z$ for different temperatures (on the right). At elevated temperatures, the QFI exhibits significant values, indicating the presence of robust entanglement among Ising spins within the QD.

As the temperature decreases, the QFI diminishes, eventually reaching zero at sufficiently low temperatures. This behavior is intriguing since, typically in a quantum spin system, lowering the temperature fosters entanglement between spins. However, in the case of the QD with Ising exchange interaction, we observe an inverse effect.

Far below the SI point, i.e., when $J_z \ll \Delta$, the spin-charge coupling is exceedingly weak, and the QFI encoded in the dynamical susceptibility is null. By increasing the exchange constant towards the $J_z/\Delta = 1$ point, we progressively approach the SI point. In this scenario, there's an escalation in dynamic transverse fluctuations of magnetization, consequently leading to an augmentation in the dynamical susceptibility and thus an increase in the entanglement within the QD. As noted earlier under the condition of $N\Delta \gg 1$, where the QD contains a considerable number of electrons, the charge and spin degrees of freedom become decoupled, and the QFI derived from the susceptibility in Eq. (13) mirrors the spin-spin entanglement in the QD.

**Dynamic transverse Susceptibility Analysis in QDs with Anisotropic Exchange Hamiltonian**

In this section, we explore the entanglement characteristics of a QD featuring an anisotropic exchange Hamiltonian ($J_z \neq J_\perp$). At elevated temperatures, when $\Delta \ll T$, employing the saddle point approximation facilitates the factorization of the QD's grand partition function into two distinct components: $Z = Z_C Z_S$. Here, $Z_C$ and $Z_S$ respectively denote the contributions of the charging energy and exchange coupling to the grand partition function.

Similar to the Ising case, the z-component of the spin operator exhibits no dynamics. However, the dynamics of the transverse components are governed by the anisotropic exchange Hamiltonian. The imaginary component of the dynamic transverse spin susceptibility is derived from $Z_S$ using the following expression:[15]

$$\mathrm{Im}\chi_{+-}(\omega) = -\frac{\pi}{2|J_z - J_\perp| Z_S} \sum_{\sigma=\pm} \left(n + \sigma T \frac{\partial}{\partial J_\perp}\right) Z_S(n)\Big|_{n=-\bar{\omega}+\sigma/2}, \quad (16)$$

where $\bar{\omega} = \omega/[2(J_z - J_\perp)]$. For the two easy-axis ($J_z \geq J_\perp$) and easy-plane ($J_z < J_\perp$) scenarios, under the condition $T \gg \Delta > J_\perp - J_z$, the partition function $Z_S$ takes the forms:[15]

$$Z_S = \begin{cases} \sqrt{\frac{\Delta}{\Delta - J_z}}\, e^{\frac{\beta J_\perp^2}{4(\Delta - J_\perp)}}\, F_1\left(\frac{\Delta}{\Delta - J_\perp}, \sqrt{\beta J_*}\right), & J_z \geq J_\perp \\ \sqrt{\frac{\Delta}{\Delta - J_z}}\, e^{\frac{\beta J_\perp^2}{4(\Delta - J_\perp)}}\, F_2\left(\frac{\Delta}{\Delta - J_\perp}, \sqrt{\beta |J_*|}\right), & J_z < J_\perp, \end{cases} \quad (17)$$

where, $J_* = \frac{(\Delta - J_\perp)(J_z - J_\perp)}{\Delta - J_z}$ represents the energy scale specific to the anisotropic problem, interpolating between 0 (for $J_z = J_\perp$) and $\frac{\Delta J_z}{\Delta - J_z}$ (for $J_\perp = 0$). The functions $F_1$ and $F_2$ are defined as follows:

$$F_1(x,y) = \int_{-\infty}^{+\infty} \frac{dt}{\sqrt{\pi}} \frac{\text{Sinh}[xyt]}{\text{Sinh}[yt]} e^{-t^2}, \; F_2(x,y) = \int_{-\frac{\pi}{2y}}^{+\frac{\pi}{2y}} \frac{dt}{\sqrt{\pi}} \frac{\text{Sin}[xyt]}{\text{Sin}[yt]} e^{-t^2}. \qquad (18)$$

In Eq. (16), the function $Z_S(n)$ is given by:[15]

$$Z_S(n) = \sqrt{\frac{\beta\Delta}{\pi}} e^{\beta(J_z - J_\perp)n^2} [\textstyle\sum_{m=|n|} e^{-\beta(\Delta - J_\perp)m^2 + \beta J_\perp m} - \sum_{m=|n|+1} e^{-\beta(\Delta - J_\perp)m^2 - \beta J_\perp m}] . \quad (19)$$

In the case where $\beta(\Delta - J_\perp)|n| \ll 1$, $Z_S(n)$ simplifies to:[15]

$$Z_S(n) = \frac{1}{2}\sqrt{\frac{\Delta}{\Delta - J_\perp}} e^{\frac{\beta J_\perp^2}{4(\Delta - J_\perp)}} e^{\beta(J_z - J_\perp)n^2} \textstyle\sum_{s=\pm} \text{erf}\,[\sqrt{\beta(\Delta - J_\perp)}(s|n| + \frac{J_\perp}{2(\Delta - J_\perp)})] + \sqrt{\frac{\beta\Delta}{\pi}} e^{-\beta(\Delta - J_z)n^2} \cosh\,[\beta J_\perp |n|], \qquad (20)$$

and in the opposite case of $\beta(\Delta - J_\perp)|n| \gg 1$, $Z_S(n)$ reduces to $\sqrt{\frac{\beta\Delta}{\pi}} e^{-\beta(\Delta - J_z)n^2 + \beta J_\perp |n|}$. Moreover, in the limit of large frequencies, i.e., for $\beta(\Delta - J_\perp)|\varpi| \gg 1$, the imaginary part of the transverse spin susceptibility is exponentially small:[15]

$$\text{Im}\chi_{+-}(\omega) = -\frac{\varpi\sqrt{\pi\beta\Delta}}{|J_z - J_\perp| Z_S} e^{-\beta(\Delta - J_z)|\varpi|(\varpi + 1) + \beta J_\perp |\varpi|}. \qquad (21)$$

In the opposite limit, when $\omega \to 0$, the imaginary part of the transverse spin susceptibility has linear behavior:

$$\text{Im}\chi_{+-}(\omega) = -\frac{\omega\sqrt{\pi\beta\Delta}}{2|J_z - J_\perp|(\Delta - J_\perp)Z_S} [\frac{2\Delta - J_\perp}{2(\Delta - J_\perp)} + \frac{\sqrt{\pi}}{2\sqrt{\beta(\Delta - J_\perp)}} g(\frac{\beta J_\perp^2}{4(\Delta - J_\perp)})], \qquad (22)$$

where $g(x) = (1 + 2x)e^x\, \text{erf}[\sqrt{x}]$.

Figure 3 presents a comparison of the behavior of $\text{Im}\chi_{+-}$ as a function of frequency (ω) for QDs with easy-axis ($J_z > J_\perp$) and easy-plane ($J_z < J_\perp$) configurations. The curve of $\text{Im}\chi_{+-}$ starts at zero for ω=0 and gradually increases at lower ω values, eventually reaching a maximum before diminishing at higher ω values.

For QDs with small exchange interactions relative to Δ (Figure 3 a and b), the responses of easy-plane and easy-axis QDs to a dynamic field are similar. As $|J_z - J_\perp|$ increases, the QDs become more anisotropic, causing their susceptibility to broaden, with the maximum shifting to higher frequencies. Unlike QDs with Ising exchange coupling, both the location and the maximum value of magnetic susceptibility in this case depend on the strengths of the coupling constants. As we approach the isotropic point ($J_z \sim J_\perp$), the susceptibility curve narrows, with a larger peak occurring at lower frequencies, eventually vanishing at the isotropic point.

When the exchange interaction is large but still smaller than Δ (Figure 3 c and d), the difference between easy-plane and easy-axis QDs becomes more pronounced. In both cases, the maximum value of susceptibility increases, but the increase in easy-plane QDs is one to two orders of magnitude greater, depending on the strengths of the exchange couplings.

In the limiting cases of $J_z = 0$ and $J_\perp = 0$ (Ising) (Figure 3 e and f), the behavior differs. In the Ising case, at a finite high temperature (e.g., $T = 5\Delta$= 58 K), the peak value does not depend on $J_z$. However, in QDs with XY exchange Hamiltonian ($J_z = 0$, and $J_\perp \neq 0$), the peak value increases with increasing $J_\perp$. For $J_\perp$ values very close to the mean level spacing $\Delta$, the $\mathrm{Im}\chi_{+-}$ peak increases sharply and tends towards infinity as $J_\perp \sim \Delta$.

For both easy-axis and easy-plane QDs with weak exchange interactions (Figure 3 a and b), the system responds to fields with lower frequencies. For instance, if the interactions are weaker than 0.1 meV, the fields required to excite the QDs fall within the range of a few tens of gigahertz, which can be readily achieved in research laboratories.

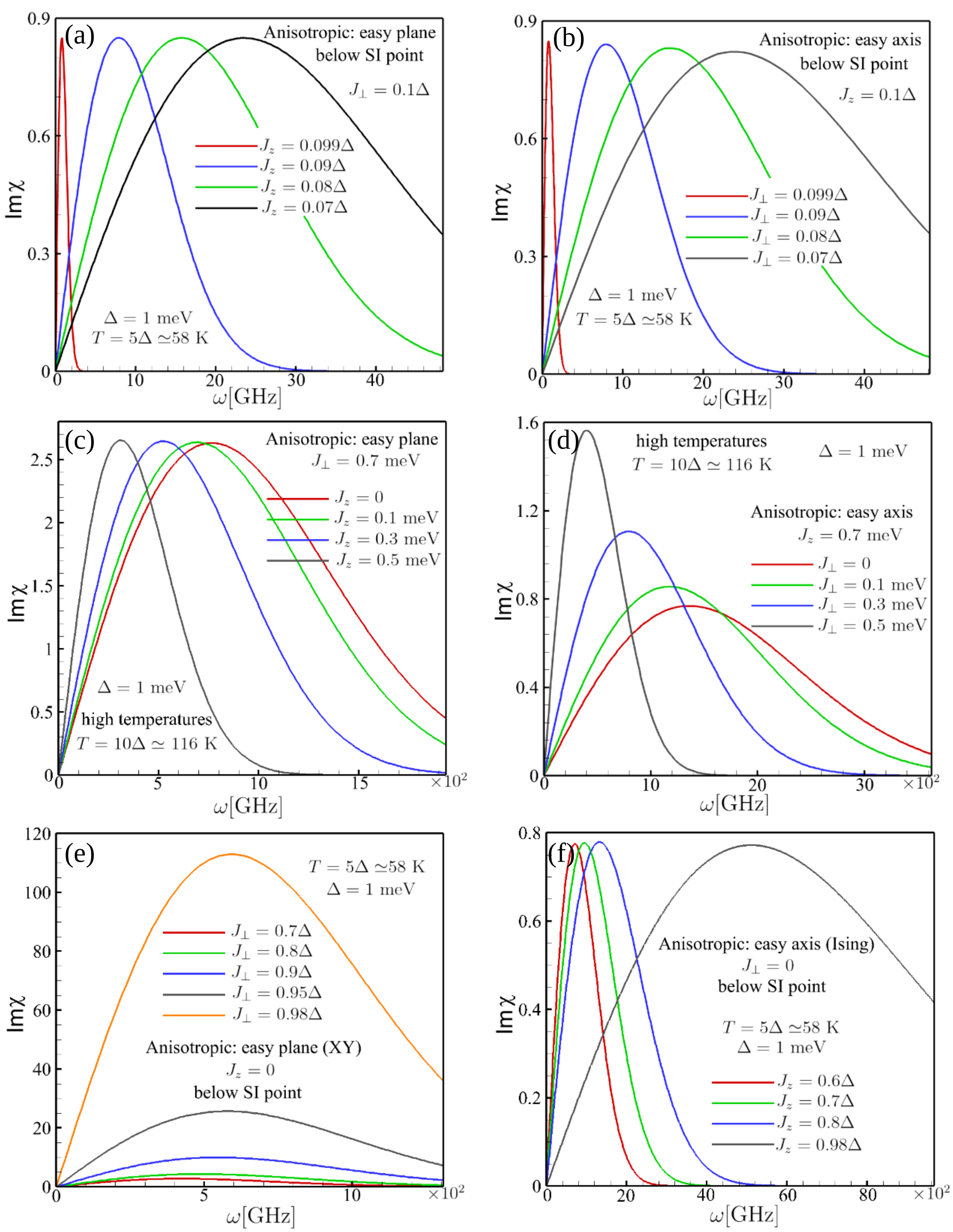

Figure 3: Plots illustrating the variation of the imaginary component of the dynamic transverse susceptibility ($\text{Im}\chi_{+-}$ ) in QDs with anisotropic exchange Hamiltonian as a function of frequency (ω). The plots compare two cases: (a, c, e) easy-plane ($J_z < J_\perp$) and (b, d, f) easy-axis ($J_z > J_\perp$) configurations. (a) and (b) depict QDs with small exchange interactions. In this scenario, the responses of the easy-plane and easy-axis QDs to a dynamic field are similar. As $|J_z - J_\perp|$ increases, the QDs become more anisotropic, causing their susceptibility to broaden, with the maximum shifting to higher frequencies. Approaching the isotropic point results in a narrower susceptibility, which eventually vanishes at the isotropic point. (c) and (d) illustrate the case where the exchange interaction is large. Here, the differences between the easy-plane and easy-axis QDs become apparent. In both cases, the maximum value of susceptibility increases, but the increase is one to two orders of magnitude greater in easy-plane QDs, depending on the strengths of the exchange couplings.

Figure 4 illustrates the variation of the imaginary component of the dynamic transverse susceptibility in QDs with an anisotropic exchange Hamiltonian as a function of frequency (ω) for various temperatures, comparing the two cases of (a, c) easy-plane (($J_z < J_\perp$) and (b, d) easy-axis ($J_z > J_\perp$) configurations. It is evident that temperature has different effects on the behavior of the dynamic response function in these QDs. For QDs with an easy axis, the maximum value of the susceptibility changes much less with increasing temperature compared to QDs with an easy plane. This difference is more significant in QDs with stronger exchange interactions as they approach Δ. In both cases, the $\text{Im}\chi_{+-}$ peak location shifts to higher frequencies as temperature increases.

To explain the results demonstrated in Figure 4, it is important to note that according to the Taylor-Maclaurin expansion for the partition function and dynamical susceptibility, the ratio of the frequency of the external perturbation to temperature plays a crucial role in the validity of this expansion. In QDs with Ising exchange, as well as in Ising-like QDs, the system responds to fields with higher frequencies as temperature increases. This causes the dynamic susceptibility to be non-zero at higher frequencies, even though the peak value of the susceptibility remains constant. This behavior is in stark contrast to QDs with an XY Hamiltonian, where the maximum value of susceptibility decreases sharply with temperature. Consequently, in Ising and Ising-like QDs, the susceptibility peak does not depend on temperature, which leads to an increase in entanglement with rising temperature (as will be discussed in the next section).

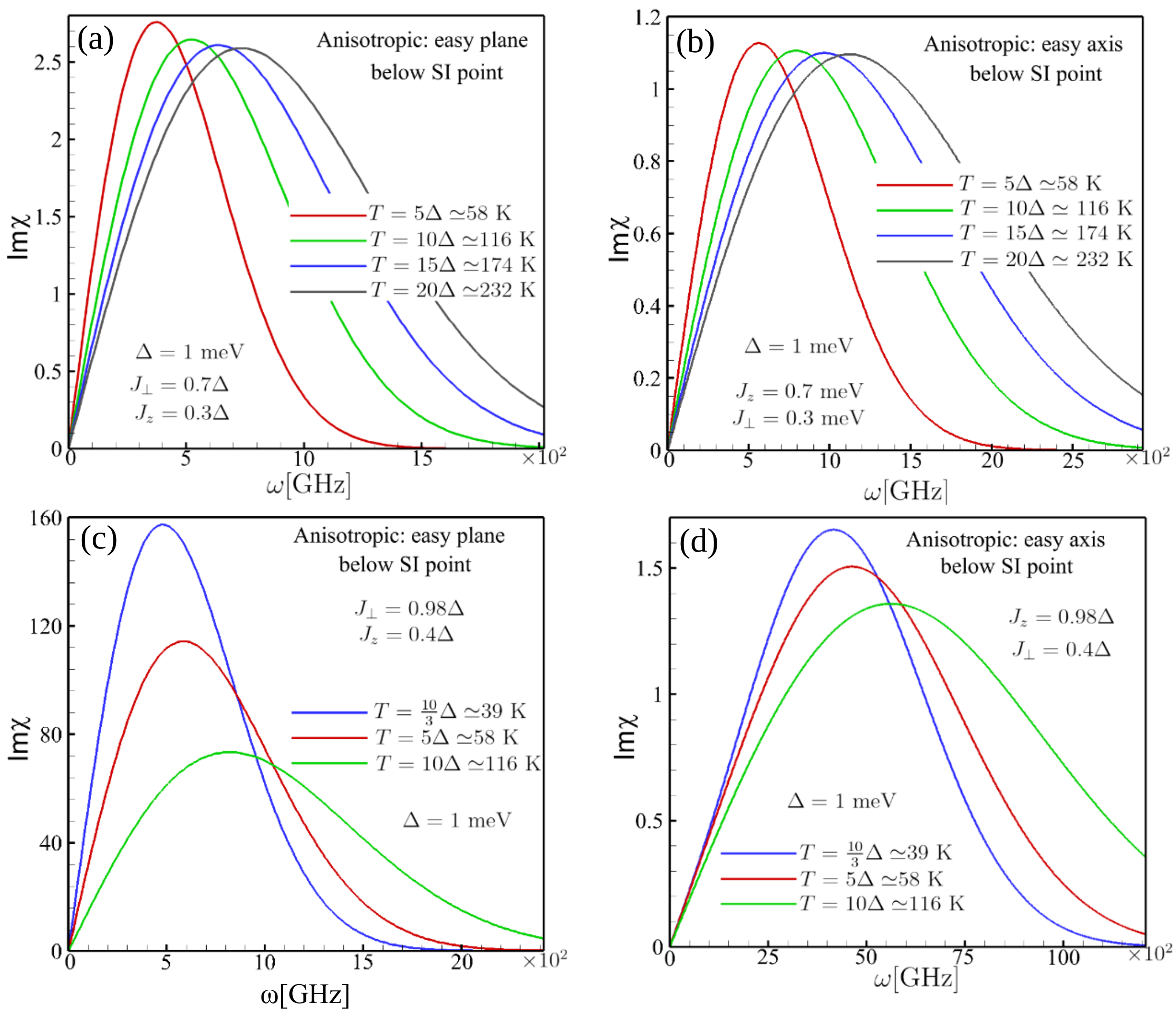

Figure 4: Plots illustrating the variation of the imaginary component of the dynamic transverse susceptibility ($\mathrm{Im}\chi_{+-}$) in QDs with anisotropic exchange Hamiltonian as a function of frequency (ω) for various temperatures. The plots compare two cases: (a, c) easy-plane ($J_z < J_\perp$) and (b, d) easy-axis ($J_z > J_\perp$) configurations. In easy-plane QDs (a, c), the peak value of $\mathrm{Im}\chi_{+-}$ increases significantly with temperature, especially when the exchange interaction is strong. In easy-axis QDs (b, d), the susceptibility's peak value changes much less with temperature. In both configurations, the peak location shifts to higher frequencies as temperature increases. This behavior is due to the ratio of the external perturbation frequency to temperature, with Ising and Ising-like QDs showing non-zero susceptibility at higher frequencies and constant peak values, unlike QDs with XY Hamiltonian where the susceptibility peak decreases with temperature.

Figure 6 demonstrates QFI as a function of temperature in QDs with an anisotropic exchange Hamiltonian for (a, b) easy-axis ($J_z > J_\perp$) and (c) easy-plane ($J_z < J_\perp$) configurations for different exchange interactions. It is evident that for both easy-axis and easy-plane QDs, QFI varies with temperature changes. In easy-plane QDs, QFI consistently decreases with increasing temperature (Figure 6c). A small QFI at high temperatures indicates that electrons in the QD are disentangled due to thermal fluctuations. By lowering the temperature, thermal fluctuations are reduced, resulting in increased spin entanglement.

This behavior contrasts with the easy-axis and Ising cases (a and b, and Figure 2a), where the temperature dependence of QFI strongly depends on the values of $J_\perp$ and $J_z$. When exchange interactions are very weak (Figure 6a), QFI increases with increasing temperature, but its value remains very small. Conversely, when exchange couplings are large (Figure 5b, $J_z = 0.9\Delta$, and different values of $J_\perp < 0.9\Delta$), QFI is large, but its temperature dependence varies with the strengths of the exchange couplings.

As clearly seen from Figure 6b, for $J_\perp \lesssim 0.43\Delta$, QFI increases with temperature, but for $J_\perp > 0.43\Delta$, QFI decreases with increasing temperature. In fact, in addition to the value of $J_z$, the parameter $J_z - J_\perp$ is also important. By increasing the value of $J_z - J_\perp$, the system approaches the Ising case, where QFI always increases with temperature.

The reason behind these differences lies in the opposite behaviors of the dynamic susceptibilities as a function of temperature. As discussed, in the latter cases, although lowering the temperature narrows the dynamic susceptibility, its peak value remains almost constant (Figure 5a and Figure 4 b, d). In contrast, in the other cases (Figure 4a, c), the peak value of the susceptibility increases significantly, leading to the enhancement of QFI with decreasing temperature.

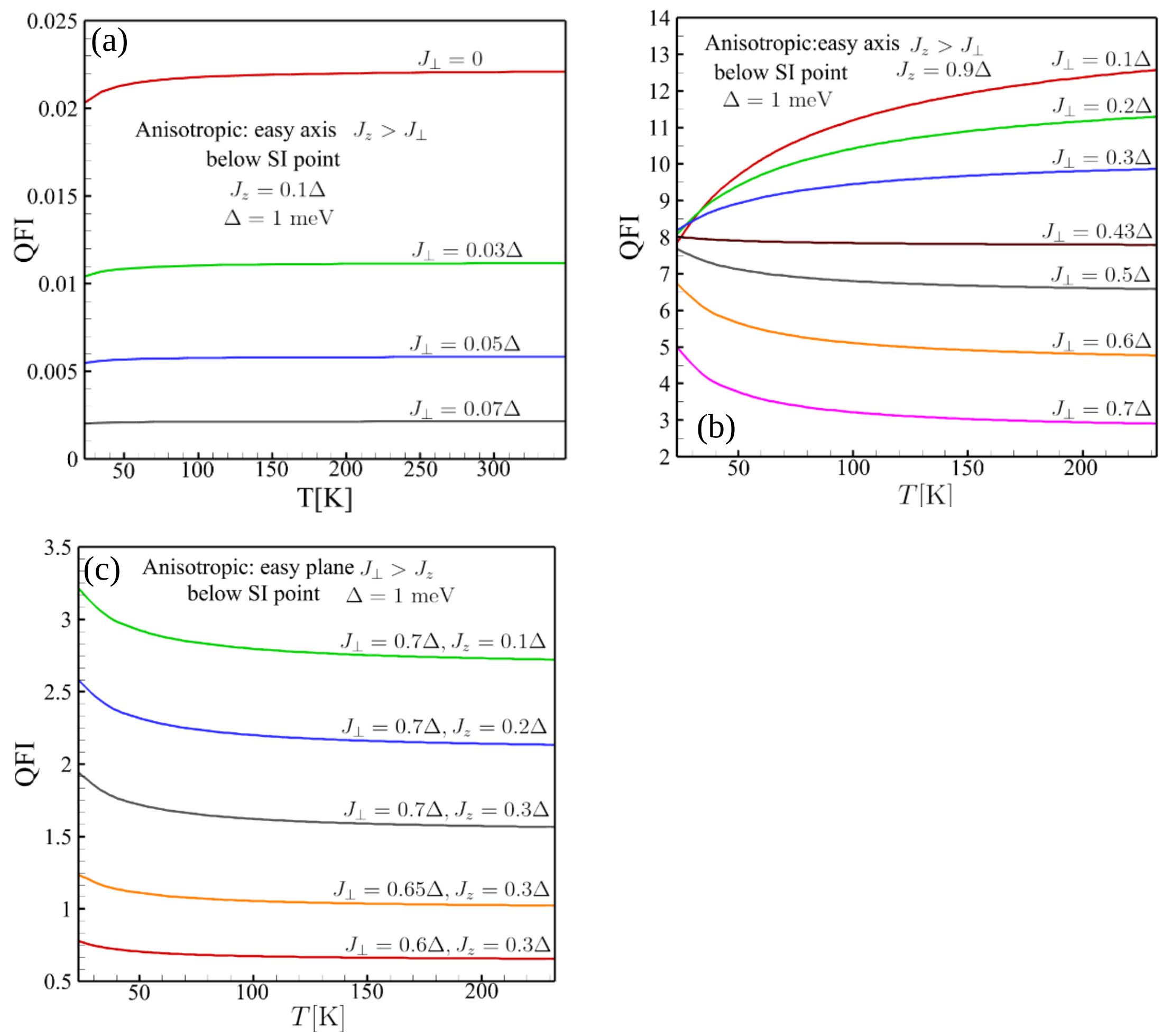

Figure 6: QFI as a function of temperature in QDs with anisotropic exchange Hamiltonian for (a, b) easy-axis ($J_z > J_\perp$) and (c) easy-plane ($J_z < J_\perp$) configurations, for different exchange interactions. (a) and (b) plot the QFI for small ($J_z = 0.1\Delta$) and large ($J_z = 0.9\Delta$) values, respectively. In QDs with an easy-plane configuration ($J_\perp > J_z$), QFI consistently decreases with increasing temperature (c). However, in the case of easy-axis QDs (a, b), the temperature behavior of QFI strongly depends on the values of $J_\perp$ and $J_z$. When exchange interactions are very weak (a), QFI increases with temperature, although its value remains very small. Conversely, when exchange couplings are large (b, $J_z = 0.9\Delta$ and various $J_\perp < 0.9\Delta$), QFI is large but its temperature dependence varies with the strength of the exchange couplings. For $J_\perp \lesssim 0.43\Delta$, QFI increases with temperature, but for $J_\perp > 0.43\Delta$, QFI decreases as temperature increases. In addition to the value of $J_z$, the parameter $J_z - J_\perp$ is also significant. As $J_z - J_\perp$ increases, the system approaches the Ising case, where QFI always increases with temperature.

Figure 7 illustrates the plot of QFI for QDs with anisotropic exchange coupling, showing its variation as a function of exchange interaction for QDs with anisotropic exchange Hamiltonian for (a) easy-axis ($J_z > J_\perp$) and (b) easy-plane ($J_z < J_\perp$) configurations, at different temperatures.

The plots of QFI with respect to the exchange couplings show that in both the easy-axis (Figure 7a) and easy-plane (Figure 7b) QDs, QFI increases when the spin Hamiltonian of the QDs is highly anisotropic. As we approach the isotropic case, the dynamic susceptibility narrows and becomes zero precisely at the isotropic point ($J_z = J_\perp$).

In the case of easy-plane QDs, the system responds more readily to the transverse magnetic field and shows relatively greater susceptibility at small frequencies compared to easy-axis QDs. This explains why QFI is larger in easy-plane QDs.

Figure 7c shows QFI for the case of easy-plane where $J_\perp$ is large (comparable to $\Delta$). In this scenario, QFI decreases with increasing $J_z$. The reduction in QFI with increasing the strength of the coupling constant at a fixed temperature is due to $J_z$ approaching $J_\perp$, the isotropic point, where the magnetic susceptibility is zero in the absence of an external magnetic field.

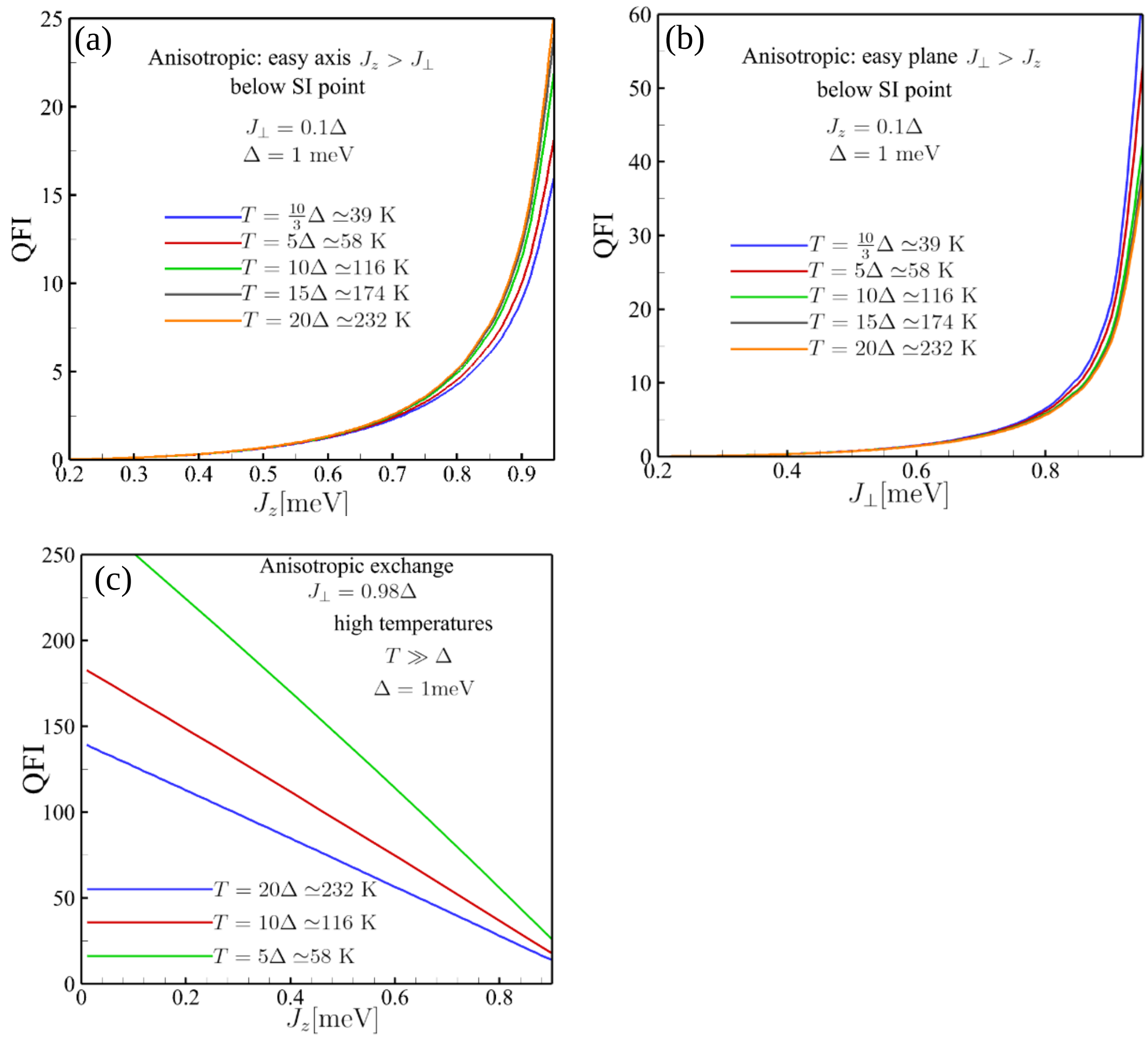

Figure 7: Plots of QFI with respect to the exchange couplings in QDs with anisotropic exchange Hamiltonian for (a) easy-axis ($J_z > J_\perp$) and (b) easy-plane ($J_z < J_\perp$) configurations at different temperatures. In both the easy-axis (a) and easy-plane (b) QDs, QFI increases when the spin Hamiltonian of the QDs is highly anisotropic. In easy-plane QDs (b), the system responds more readily to the transverse magnetic field, exhibiting greater susceptibility at small frequencies compared to easy-axis QDs (a). (c) shows the case of easy-plane QDs where $J_\perp$ is large (comparable to $\Delta$). In this scenario, QFI decreases with increasing $J_z$ .

In the above calculations, it should be noted that we have neglected the influence of QD's level fluctuations on the behavior of susceptibility and entanglement. We anticipate that the interplay between level fluctuations and exchange interactions will give rise to intriguing physics related to the quantum entanglement of the QD.

## Spin-Spin Entanglement in QDs at Low Temperatures

In the low-temperature regime ($T \ll \Delta$), the behavior of QDs with anisotropic spin Hamiltonian can be effectively described using the grand partition function expressed as:[15]

$$Z = \sum_{n_\uparrow n_\downarrow} Z_{n_\uparrow} Z_{n_\downarrow} e^{-\beta E_c (n-N_0)^2+\beta\mu n+\beta J_\perp m(m+1)} \mathrm{Sgn}(2m+1) \sum_{l=-|m+1/2|+1/2}^{|m+1/2|-1/2} e^{\beta(J_z-J_\perp)l^2}, \quad (23)$$

where $n = n_\uparrow + n_\downarrow$ and $m = (n_\uparrow - n_\downarrow)/2$. The function $Z_n$ is the canonical partition function of the QD with $n$ number of electrons. The dynamical susceptibility is given by:

$$\chi_{\pm}(\omega) = \frac{1}{Z} \Sigma_{n_\uparrow, n_\downarrow} Z_{n_\uparrow} Z_{n_\downarrow} e^{-\beta E_c (n-N_0)^2 + \beta \mu n + \beta J_\perp m(m+1)} \operatorname{sgn}(2m+1) \Sigma_{l=-\left|m+\frac{1}{2}\right|+\frac{1}{2}}^{\left|m+\frac{1}{2}\right|-\frac{1}{2}}$$

$$e^{\beta (J_z - J_\perp) l^2} \left[ \frac{m(m+1) - l^2 - l}{\omega + (J_\perp - J_z)(2l+1) + i0^+} - \frac{m(m+1) - l^2 + l}{\omega + (J_\perp - J_z)(2l-1) + i0^+} \right] \quad (24)$$

The imaginary part of susceptibility is:

$$\mathrm{Im}\chi_{\pm}(\omega) = \frac{-\pi}{Z} \Sigma_{n_\uparrow, n_\downarrow} Z_{n_\uparrow} Z_{n_\downarrow} e^{-\beta E_c (n-N_0)^2 + \beta \mu n + \beta J_\perp m(m+1)} \operatorname{sgn}(2m+1) \Sigma_{l=-\left|m+\frac{1}{2}\right|+\frac{1}{2}}^{\left|m+\frac{1}{2}\right|-\frac{1}{2}}$$

$$e^{\beta (J_z - J_\perp) l^2} [(m(m+1) - l^2 - l) \delta(\omega + (J_\perp - J_z)(2l+1))$$

$$-(m(m+1) - l^2 + l) \delta(\omega + (J_\perp - J_z)(2l-1))] \quad (25)$$

In the regime of low temperatures ($T \ll \Delta$), where thermal fluctuations are minimal, the canonical partition function can be approximated as $Z_n \sim e^{-\beta \Delta n(n-1)/2}$. In Equation (25), we must consider the contribution where $m = S$ and $l = \pm S$ only, where $S$ represents the total spin of the ground state.

In this limit, the imaginary part of the dynamical susceptibility simplifies to a series of Dirac delta functions, given by:

$$\mathrm{Im}\chi_{+-}(\omega) = \pi S [\delta(\omega - (J_z - J_\perp)(2S-1)) - \delta(\omega + (J_z - J_\perp)(2S-1))] \quad (26)$$

This expression reflects the behavior of the system's susceptibility at low temperatures, indicating the discrete nature of the spin transitions within the QD.

The behavior of $\mathrm{Im}\chi$ strongly relies on the value of the total spin $S$. As discussed earlier, within the mesoscopic Stoner regime where the conditions $\Delta - J_z \ll J_\perp < J_z$ are satisfied, the total spin of the ground state can be approximated as $S \cong (J_\perp + J_z - \Delta)/[2(\Delta - J_z)]$. This expression yields a positive variable for $S$. In terms of the mean level spacing, the second term in the argument of the delta functions simplifies to $(J_z - J_\perp)\left(\frac{J_\perp}{1-J_z} - 2\right)$, which is always positive within the mesoscopic Stoner regime.

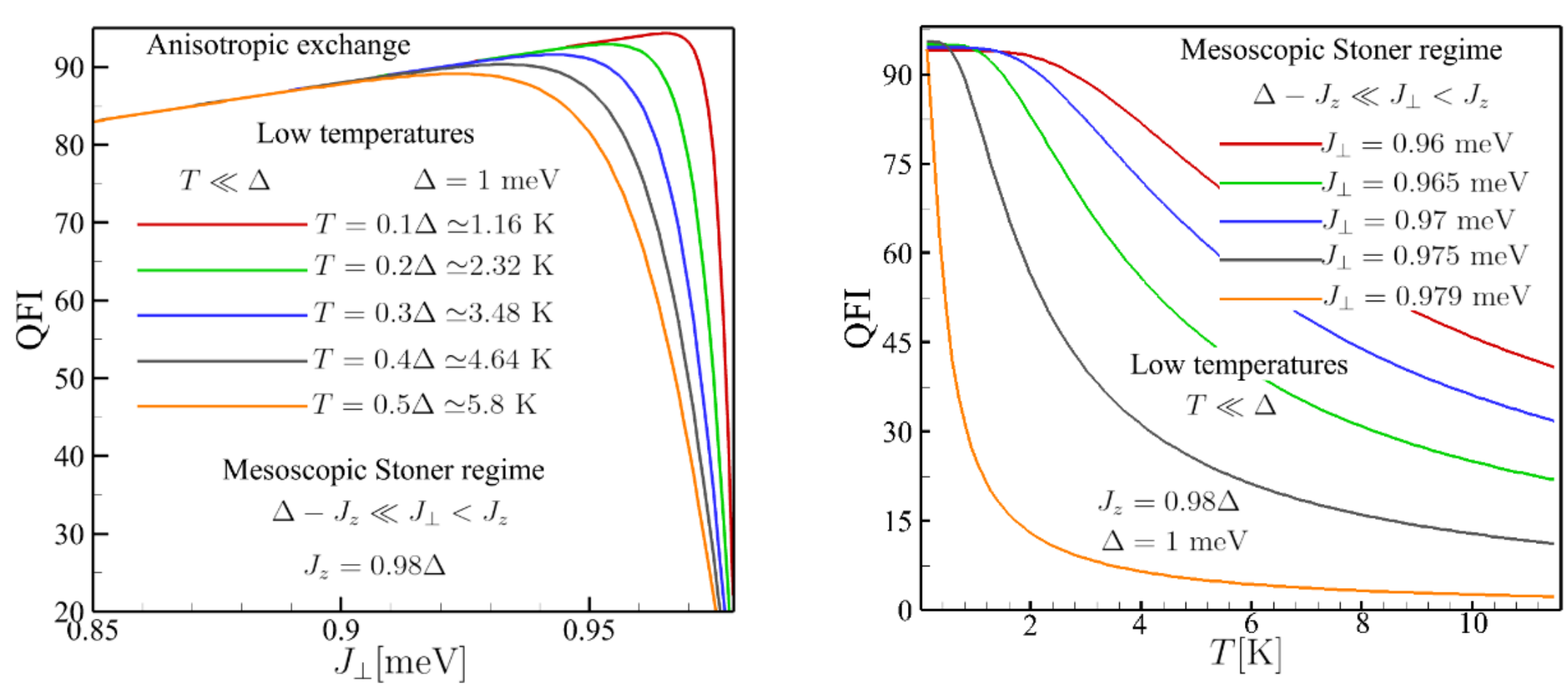

Figure 8: QFI plotted as a function of $J_\perp$ in QDs with anisotropic exchange Hamiltonian for various temperatures (left), and as a function of temperature for $J_z > J_\perp$ (right).

Substituting Eq. (26) into Eq. (15), the QFI is obtained as $\text{QFI} = 4S \tanh\left[\frac{1}{2T}(J_z - J_\perp)(2S - 1)\right]$. At sufficiently low temperatures, the hyperbolic tangent function approaches 1, and the QFI simplifies to $4S$. For a given value of $\Delta$ and $J_z$, the ground state spin increases with increasing $J_\perp$, leading to an increase in QFI (refer to Figure 8-left). In contrast to the high-temperature limit (see Figure 7-right for comparison), here the QFI rises as the system approaches the isotropic point, peaks near this point, and then sharply decreases to zero at $J_z = J_\perp$. At this juncture, where the transverse spin susceptibility is zero, the QFI also becomes zero, consistent with its behavior at high temperatures.

In the Ising model case ($J_\perp = 0$), as long as $\Delta > J_z$, the ground state spin is $S = \frac{1}{2}$ (0) for an odd (even) number of electrons. In both scenarios, the QFI evaluates to zero.

**Practical Implications and Relevance of the Study**

The parameters in the plots have been intentionally scaled to measurable units, with an assumption of an energy level spacing of 1 meV, corresponding to a planar QD or nanoparticle dimension of approximately 100 nm in diameter. This allows the reported plots to reflect meaningful values for experimentalists. For example, the dynamic susceptibility spectra, plotted versus frequency, show that the peaks occur in the range of 10-500 GHz. While this frequency range may pose challenges for measurement in practice, it is worth noting that under applied magnetic fields, this range can be reduced. Moreover, the metallic domain assumption is applicable to many-electron QDs, nanoparticles, or metallic impurities in a host material. These practical observations provide valuable insights for experimentalists seeking to validate the theoretical predictions in real-world scenarios.

The findings of this study hold practical implications in several key areas of quantum research and technology. Firstly, by elucidating the behavior of QDs under various types of exchange interaction, the study contributes to the development of more efficient means of determining the entanglement in QDs. Understanding the impact of diverse exchange interactions within QDs is paramount for designing QDs with tailored phase stability or coherence properties.

QDs serve as exceptional platforms for exploring quantum effects at the nanoscale, providing unparalleled control over individual quantum states. Entanglement, a fundamental resource for quantum technologies, highlights the system's sensitivity to environmental factors. In this regard, by demonstrating dynamic susceptibility as a proxy to detect and manipulate entanglement in QDs, this study paves the way for the development of high-precision quantum sensors and measurement devices. Such advancements could find applications in fields like magnetometry, navigation, and medical imaging, offering improved sensitivity and accuracy in quantum metrology tools.

Moreover, the study's emphasis on addressing practical challenges posed by stoner instability underscores the necessity of considering these factors in the design and fabrication of quantum devices. By elucidating the influence of various types of exchange interactions on QD behavior, this research informs the development of robust and reliable quantum devices capable of functioning effectively in real-world conditions.

**Future Directions: Quantum Metrology and Beyond with AC Susceptibility Measurements**

The proposed characterization method offers scalability advantages compared to resource-intensive techniques outlined in references 15 and 18. While traditional methodologies excel in accessing non-local entanglement, a hallmark of topological phases as discussed in references 16 and 17, physical susceptibilities establish closer ties with local operators.

Despite QFI displaying scaling behavior at topological transitions, it falls short of identifying topological phases, as demonstrated in [1] using the Kitaev wire example. This limitation is shared by other entanglement witnesses.[18,19]

A key challenge lies in exploring non-local extensions of the proposed protocol, potentially enabling the characterization of topological entanglement[20] and exotic quantum phases.[21,22]

The multipartite entanglement detected by QFI serves as a valuable resource for quantum metrology.[16,17] This could expand its application scope to include quantification of entanglement in quantum simulators for many-body problems[23,24,25] and characterization of strongly correlated systems.

For example, one intriguing avenue to explore is whether entanglement influences material properties such as high-temperature superconductivity, where scaling behavior has already been observed in optical conductivity.[26] Additionally, an underlying antiferromagnetic critical point[27] is expected to exhibit quantum-critical scaling of entanglement, a theory that could be tested in cold atomic-gas experiments.[2,3]

In both the isotropic and anisotropic QDs we investigated, the null dynamic longitudinal susceptibility indicates that the z-component of the spin operators in the QD remains constant over time. Consequently, the electron-electron interaction does not influence the longitudinal relaxation time ($T_1$). Conversely, the transverse components of the spin operators exhibit correlations at different times, resulting in a nonzero dynamic transverse susceptibility and finite entanglement between electrons. This entanglement plays a crucial role in determining the transverse relaxation time ($T_2$). Therefore, to design a QD with long relaxation times, efforts should focus on minimizing the entanglement of particles induced by the exchange interaction. Notably, these results were obtained for QDs containing a large number of electrons. In the context of qubit QDs, reducing the number of electrons within the QD becomes essential. It remains an intriguing question whether the transverse relaxation time is still dependent on the exchange interaction of particles in such scenarios.

**Conclusion**

This study examines the entanglement in QDs, shedding light on various phenomena influenced by confinement, exchange interactions, temperature variations, and proximity to the SI point. Our investigation reveals that magnetization in QDs is significantly influenced by their proximity to the SI point. Charge-spin coupling is strong near the SI point but weakens as it moves away, with magnetization fluctuations near the SI point strengthening charge-spin interactions. Furthermore, our analysis shows that the canonical partition function and Coulomb blockade regime play crucial roles in governing electron count and charging in QDs. Far from the SI point, charge and spin become decoupled, as evidenced by magnetic susceptibilities showing no charge interaction.

By comparing the behaviors of the imaginary component of the dynamic transverse susceptibility ($\mathrm{Im}\chi_{+-}$) for QDs with easy-axis ($J_z > J_\perp$) and easy-plane ($J_z < J_\perp$) configurations, several key insights emerge:

1. For QDs with small exchange interactions, the responses of easy-plane and easy-axis QDs to a dynamic field are similar. As $|J_z - J_\perp|$ increases, the QDs become more anisotropic, their susceptibility broadens, and its maximum forms at higher frequencies. Near the isotropic point, the susceptibility narrows and finally vanishes.

2. When the exchange interaction is large, differences between easy-plane and easy-axis QDs become apparent. The maximum value of the susceptibility increases in both cases, but the increase in easy-plane QDs is one to two orders of magnitude greater, depending on the strength of the exchange couplings.
3. In the Ising case ($J_\perp = 0$, $J_z \neq 0$), the QD responds to dynamic magnetic fields at higher frequencies, with $\mathrm{Im}\chi_{+-}$ peaking at higher frequencies.
4. At high temperatures (e.g., T=5Δ=58 K), the peak value of $\mathrm{Im}\chi_{+-}$ in Ising QDs does not depend on $J_z$. However, in QDs with XY exchange Hamiltonian ($J_\perp \neq 0, J_z = 0$), the peak value increases with increasing $J_\perp$ and tends to infinity as $J_\perp \approx \Delta$.
5. For both easy-axis and easy-plane QDs with weak exchange interactions, the system responds to fields at lower frequencies. For example, with interactions weaker than 0.1 meV, the required fields to excite QDs are in the range of a few tens of gigahertz, achievable in research laboratories.
6. Temperature effects on the dynamic response function differ between these QDs. In easy-axis QDs, the maximum value of susceptibility remains unchanged with increasing temperature, but the peak location shifts to higher frequencies. Conversely, in easy-plane QDs, the maximum value decreases, and the response function broadens with increasing temperature, especially with stronger exchange interactions.
7. According to the Taylor-Maclaurin expansion for the partition function and dynamical susceptibility, the ratio of the frequency of external perturbation to temperature is crucial. In Ising and Ising-like QDs, as temperature increases, the system responds to higher frequency fields, causing non-zero dynamical susceptibility at higher frequencies. This leads to increased entanglement with rising temperature, unlike in QDs with XY Hamiltonian where the susceptibility peak decreases with temperature.

The study also reveals that QFI varies with temperature for both easy-axis and easy-plane QDs. In easy-plane QDs ($J_z < J_\perp$), QFI consistently decreases with increasing temperature. In contrast, the temperature behavior of QFI in easy-axis QDs depends on the values of $J_\perp$ and $J_z$. For weak exchange interactions, QFI increases with temperature, but its value is very small. For large exchange couplings ($J_z = 0.9\Delta$ and $J_\perp < 0.9\Delta$), QFI is large but varies with the strengths of the exchange couplings. Specifically, for $J_\perp \lesssim 0.43\Delta$, QFI increases with temperature, whereas for $J_\perp > 0.43\Delta$, QFI decreases with temperature. The parameter $J_z - J_\perp$ is crucial, as increasing it approaches the Ising case, where QFI always increases with temperature.

The plots of QFI with respect to the exchange couplings show that in both easy-axis and easy-plane QDs, QFI increases when the spin Hamiltonian is highly anisotropic. Approaching the isotropic case, the dynamic susceptibility narrows, causing QFI to decrease. Easy-plane QDs respond more easily to transverse magnetic fields, exhibiting greater susceptibility at small frequencies, which explains the larger QFI in easy-plane QDs.

These findings deepen our understanding of entanglement in QDs and have practical implications for quantum research and technology. By elucidating the intricate interplay between exchange interactions, temperature variations, and entanglement, this study lays the groundwork for future investigations aimed at harnessing entanglement for quantum applications.

## Acknowledgement

This study is partially based on work supported by AFOSR and LPS under contract numbers FA9550-23-1-0302 and FA9550-23-1-0763, and by the NSF under grant number CBET-2110603.

[18] Amico, L., Fazio, R., Osterloh, A., and Vedral, V., “Entanglement in many-body systems.” Rev. Mod. Phys., 80 517 (2008).
[19] Gühne, O., and Tóth, G., “Entanglement detection.” Phys. Rep., 474 1 (2009).
[20] Zeng, B., Chen, X., Zhou, D.-L., and Wen, X.-G., “Quantum information meets quantum matter from quantum entanglement to topological phase in many-body systems.” Preprint at http://arxiv.org/abs/1508.02595 (2015).
[21] Balents, L., “Spin liquids in frustrated magnets.” Nature, 464 199 (2010).
[22] Han, T.-H. et al. “Fractionalized excitations in the spin-liquid state of a kagome-lattice antiferromagnet.” Nature, 492 406 (2012).
[23] Cirac, J. I. & Zoller, P., “Goals and opportunities in quantum simulation.” Nature Phys., 8 264 (2012).
[24] Hauke, P., Cucchietti, F. M., Tagliacozzo, L., Deutsch, I., and Lewenstein, M., “Can one trust quantum simulators?” Rep. Prog. Phys., 75 082401 (2012).
[25] Georgescu, I. M., Ashhab, S. & Nori, F., “Quantum simulation.” Rev. Mod. Phys., 86 153 (2014).
[26] van der Marel, D., et al., “Quantum critical behavior in a high-Tc superconductor.” Nature, 425 271 (2003).
[27] Dai, P., “Antiferromagnetic order and spin dynamics in iron-based superconductors.” Rev. Mod. Phys., 87 855 (2015).